\documentclass{iopart}

\usepackage{graphicx, dblfloatfix}
\usepackage{iopams}
\usepackage{hyperref}
\usepackage{cite}
\usepackage{color, xcolor}
\usepackage{rotating}
\usepackage{tabularx}

\usepackage{booktabs}
\usepackage{multirow}
\usepackage{makecell}   
\usepackage{siunitx}    

\usepackage{amssymb}
\usepackage{braket}
\usepackage{verbatim}
\usepackage{verbatim}
\usepackage{float}

\expandafter\let\csname equation*\endcsname\relax
\expandafter\let\csname endequation*\endcsname\relax
\usepackage{amsmath}

\newcommand{\trm}{\textrm}

\newcommand{\corr}[1]{\textcolor{red}{#1}}

\usepackage{lineno}
\usepackage[T1]{fontenc}

\begin{document}
\title[]{Control of pedestal-top electron density using RMP and gas puff at KSTAR
\\
}

\date{\today}

\author{Minseok~Kim$^{1\dagger}$, S.K.Kim$^2$, A.Rothstein$^1$, P.Steiner$^1$, K.Erickson$^2$, Y.H.Lee$^3$, H.Han$^3$, Sang-hee~Hahn$^3$, J.W.Juhn$^3$, B.Kim$^{3,5}$, R.Shousha$^{2}$, C.S.Byun$^1$, J.Butt$^1$, ChangMin~Shin$^4$, J.Hwang$^{3,4}$, Minsoo Cha$^5$, Hiro Farre$^1$, S.M.Yang$^2$, Q.Hu$^2$, D.Eldon$^6$, N.C.Logan$^7$, A.Jalalvand$^1$, and E.Kolemen$^{1\ast}$\footnote{Author to whom any correspondence should be addressed.}}

\address{$^1$Department of Mechanical and Aerospace Engineering, Princeton University, Princeton, NJ 08540, USA}
\address{$^2$Princeton Plasma Physics Laboratory, Princeton, NJ, 08540, USA}
\address{$^3$Korea Institute of Fusion Energy (KFE), Daejeon 34133, Republic of Korea}
\address{$^4$Department of Nuclear and Quantum Engineering, Korea Advanced Institute of Science and Technology, Daejeon 34141, Republic of Korea}
\address{$^5$Department of Nuclear Engineering, Seoul National University, Seoul, 08826, Republic of Korea}
\address{$^6$General Atomics, San Diego, CA, 92121, USA}
\address{$^7$Department of Applied Physics, Columbia University, New York, NY, 10027, USA}

\ead{\mailto{$^\dagger$mseokim@princeton.edu}, \mailto{$^\ast$ekolemen@princeton.edu}}


\begin{abstract}
We report the experimental results of controlling the pedestal-top electron density by applying resonant magnetic perturbation with the in-vessel control coils and the main gas puff in the 2024-2025 KSTAR experimental campaign. The density is reconstructed using a parametrized $\psi_\trm{N}$ grid and the five channels of the line-averaged density measured by a two-colored interferometer. The reconstruction procedure is accelerated by deploying a multi-layer perceptron to run in about 120 $\mu s$ and is fast enough for real-time control. A proportional-integration controller is adopted, with the controller gains being estimated from the system identification processes. The experimental results show that the developed controller can follow a dynamic target while exclusively using both actuators. The absolute percentage errors between the electron density at $\psi_\trm{N}=0.89$ and the target are approximately \SI{1.5}{\percent} median and a \SI{2.5}{\percent} average value. The developed controller can even lower the density by using the pump-out mechanism under RMP, and it can follow a more dynamic target than a single actuator controller. The developed controller will enable experimental scenario exploration within a shot by dynamically setting the density target or maintaining a constant electron density within a discharge.
\end{abstract}

\maketitle
\ioptwocol

\section{Introduction}
\label{sec:intro}

Research in tokamaks has led the nuclear fusion society with its higher performance than other competitors since the high confinement mode (H-mode) \cite{paper:wagner1982regime} was discovered. H-mode operation is desirable for future nuclear fusion reactors due to its higher plasma confinement, but the burst of particles and heat at the edge region, which is called edge localized mode (ELM) \cite{paper:wagner1984importance, paper:gohil1988study, paper:zohm1992studies, paper:evans, paper:kim2020nonlinear, paper:kim2020pedestal, paper:yang2024tailoring, paper:kim2024highest}, would occur under the operational regime. A single ELM would crack the wall for the reactor scale devices, so an ELM-free state is necessary. To mitigate the issue, there have been trials to control the ELMs under resonant magnetic perturbation (RMP) from in-vessel control coils (IVCC) \cite{paper:shousha2022design, paper:kim2022optimization, paper:shousha2023thesis}.

In addition to the ELM-free operation, detachment \cite{paper:matthews1995plasma, paper:loarte1998plasma, paper:loarte2001effects} is another necessary operational regime for the long-pulse steady state plasmas, which would put less heat load on the divertor plate. Therefore, ELM-free and detached long-pulse steady state plasmas with H-mode \cite{abstract:hu2023integration} are needed to run a tokamak nuclear fusion power plant. Since the electron density ($n_\mathrm{e}$) at the pedestal region is one of the key factors affecting ELM and detachment, the feasible window of the $n_\mathrm{e}$ needs to be explored to achieve either state. Additionally, $n_\mathrm{e}$ should be kept constant throughout the operation to ensure it is in a steady state. This motivates us to develop a pedestal-top electron density controller, which can explore the density level that allows the ELM-free or detached state and keep the density level constant throughout the operation. There have been attempts to control pedestal-top $n_\textrm{e}$ with ELM suppression \cite{paper:laggner2020real} and with $W_\textrm{MHD}$ \cite{paper:hawryluk2015control} using both resonant magnetic perturbation (RMP) and gas puff. Inspired by previous works, we present experimental results at KSTAR, where we control the pedestal-top $n_e$ to follow a dynamic target by mutually exclusively using both RMP from IVCC and the Piezo-electric Valve Midplane G-port (PVMG) with $D_2$, in short, a main gas puff in a feedback manner.

The $n_\mathrm{e}$ profile has been reconstructed offline at KSTAR by leveraging the information from the equilibrium reconstruction algorithm (EFIT) \cite{paper:EFIT, paper:CAKE, paper:Li2013, paper:jiang2021} and the line-averaged density information measured by the two-colored interferometer (TCI) system \cite{paper:juhn2021multi}. The profile reconstruction procedure is to find the parametrized $n_\mathrm{e}$ profile consistent with the TCI measurements on the $\psi_{\mathrm{N}}$ grid reconstructed by EFIT. Even though the procedure has been routinely used for the post-experiment analysis, it was not fast enough for the KSTAR plasma control system (PCS) for real-time analysis and control purposes. Accordingly, it is necessary to accelerate the profile fitting routine to monitor and control the values in $n_\textrm{e}$ profile in real-time. Previously, the acceleration of reconstruction or prediction of physical quantities has been done in nuclear fusion research by adopting neural network models \cite{paper:shousha2023machine, paper:rothstein2025torbeamnn, paper:boyer2021prediction, paper:joung2019deepNN, paper:joung2023gs}. Inspired by the previous work, we also implemented a simple multilayer perceptron (MLP) \cite{book:bishop} neural network to accelerate the process, which takes the five quantities parameterizing the $\psi_\textrm{N}$ grid and the measurements from five TCI channels as inputs (ten in total) and the four fitting parameters as outputs. The neural network has been implemented in the KSTAR PCS.

The $n_\textrm{e}$ value at $\psi_\textrm{N} = 0.89$ which is the position around the pedestal-top was controlled by deploying a proportional-integration (PI) controller \cite{book:Astrom, book:Rowley} whose inputs are the currents of IVCC \cite{paper:kim2009IVCCdesign, paper:kim2011IVCCfabrication} and voltage applied to the main gas puff \cite{kim2013gas}. The currents of IVCC can apply RMP to the plasmas, which can suppress ELM and pump out the density around the pedestal-top region, and the voltage of the main gas puff can inject the main gas, $D_2$, into the plasmas. The controller was designed by analyzing the plasma response to the actuators, and it was tested in multiple experimental sessions in the 2024-2025 KSTAR campaign.

The rest of the paper is organized as follows: Section. \ref{sec:nerecon} describes the real-time $n_\textrm{e}$ profile reconstruction procedure deploying the MLP. The fitting model used for the reconstruction, the reconstruction procedure exploiting the information from TCI and EFIT, and the acceleration process adopting an MLP are included. Subsequently, the real-time control of the $n_\textrm{e}$ at $\psi_\textrm{N}=0.89$ implementing the PI controller is described in the Section. \ref{sec:necontrol} where the system identification procedure to design the controller is illustrated, the detailed controller design procedure is specified, and the experimental results are shown. The paper is summarized, and future work is suggested in Sec. \ref{sec:conclusion}.

\section{Electron density profile reconstruction}
\label{sec:nerecon}

\begin{figure*}[htbp]
\begin{center}
\includegraphics[width=\linewidth]{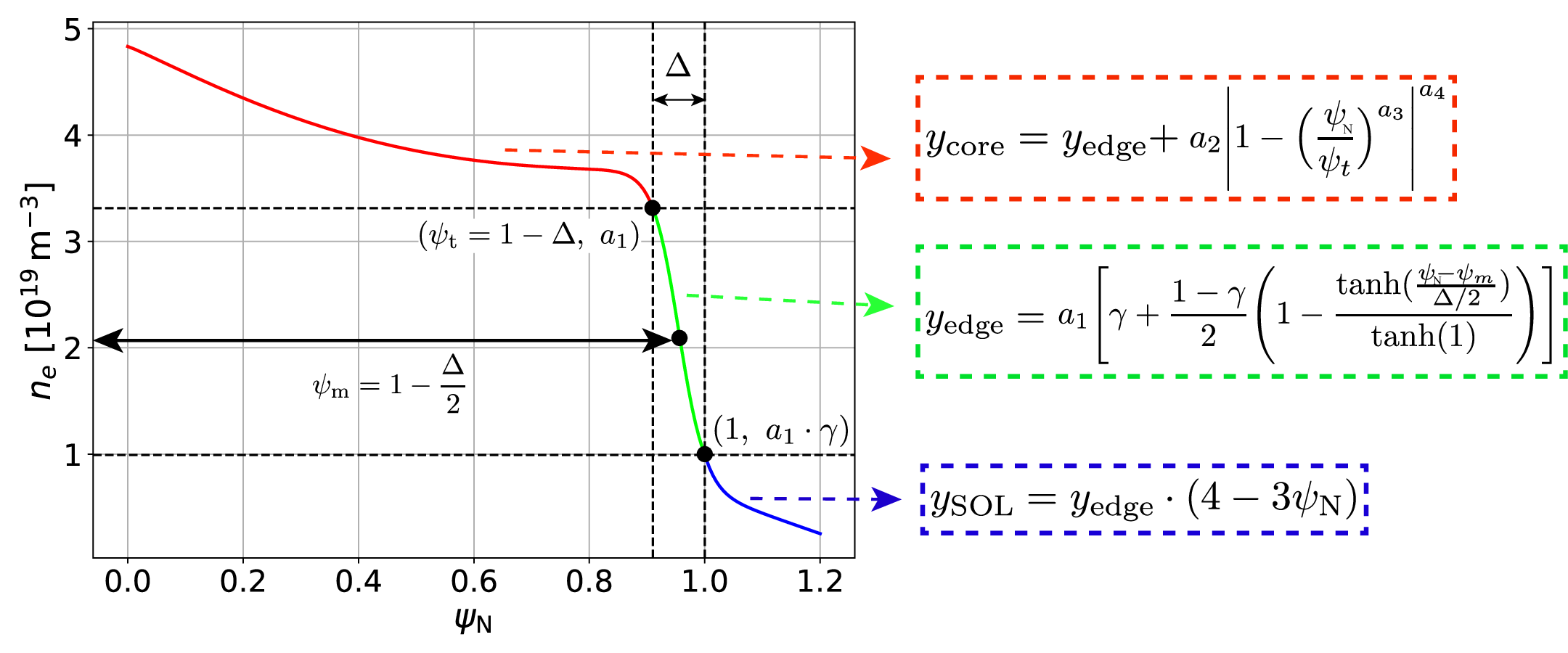}
\caption[]{Description of the fitting function to reconstruct the $n_\textrm{e}$ profile. The edge model expresses the pedestal structure with a hyperbolic tangent function, the core model is a polynomial function to let it be flexible, and a linear function expresses the scrape-off layer.}
\label{fig:fitting}
\end{center}
\end{figure*}

\subsection{Fitting model}
\label{subsec:model}

To get spatially continuous kinetic profiles inside a tokamak, including toroidal velocity, ion and electron temperature, and electron density, out of spatially discrete measurements, we need a profile fitting scheme. This can be achieved by non-parametric or parametric fitting functions\cite{paper:fitting}. Non-parametric fitting methods, such as Gaussian process regression\cite{book:Rasmussen, paper:ms_GPR}, are more flexible than their competitors, allow us to express our prior beliefs during the fitting procedure, and are easy to use to infer a physical quantity consistent with numerous diagnostics\cite{paper:kwak}. While Bayesian methods are superior for incorporating prior information, they are too computationally expensive. Therefore, we decided to use a parametric fitting method, which is computationally faster than non-parametric ones, so that it can be applied for real-time control.

The parametric fitting function we used is described in Fig. \ref{fig:fitting}. We assume that the edge profile would follow a tangent-hyperbolic-like shape, and the core would follow a polynomial shape. The scrape-off layer is expressed by using a simple linear function. There are six parameters of the fitting function, $a_1$, $a_2$, $a_3$, $a_4$, $\Delta$, and $\gamma$. 

\begin{itemize}
  \item $a_1$: $n_\textrm{e}$ at the pedestal-top
  \item $a_2$: Difference of the $n_\textrm{e}$ between the magnetic axis and pedestal-top
  \item $a_3$: Fitting coefficient in the core region
  \item $a_4$: Fitting coefficient in the core region
  \item $\Delta$: The pedestal width
  \item $\gamma$: The ratio of $n_\textrm{e}$ between pedestal-top and the last closed flux surface (LCFS)
\end{itemize}

\begin{figure*}[htbp]
\begin{center}
\includegraphics[width=\linewidth]{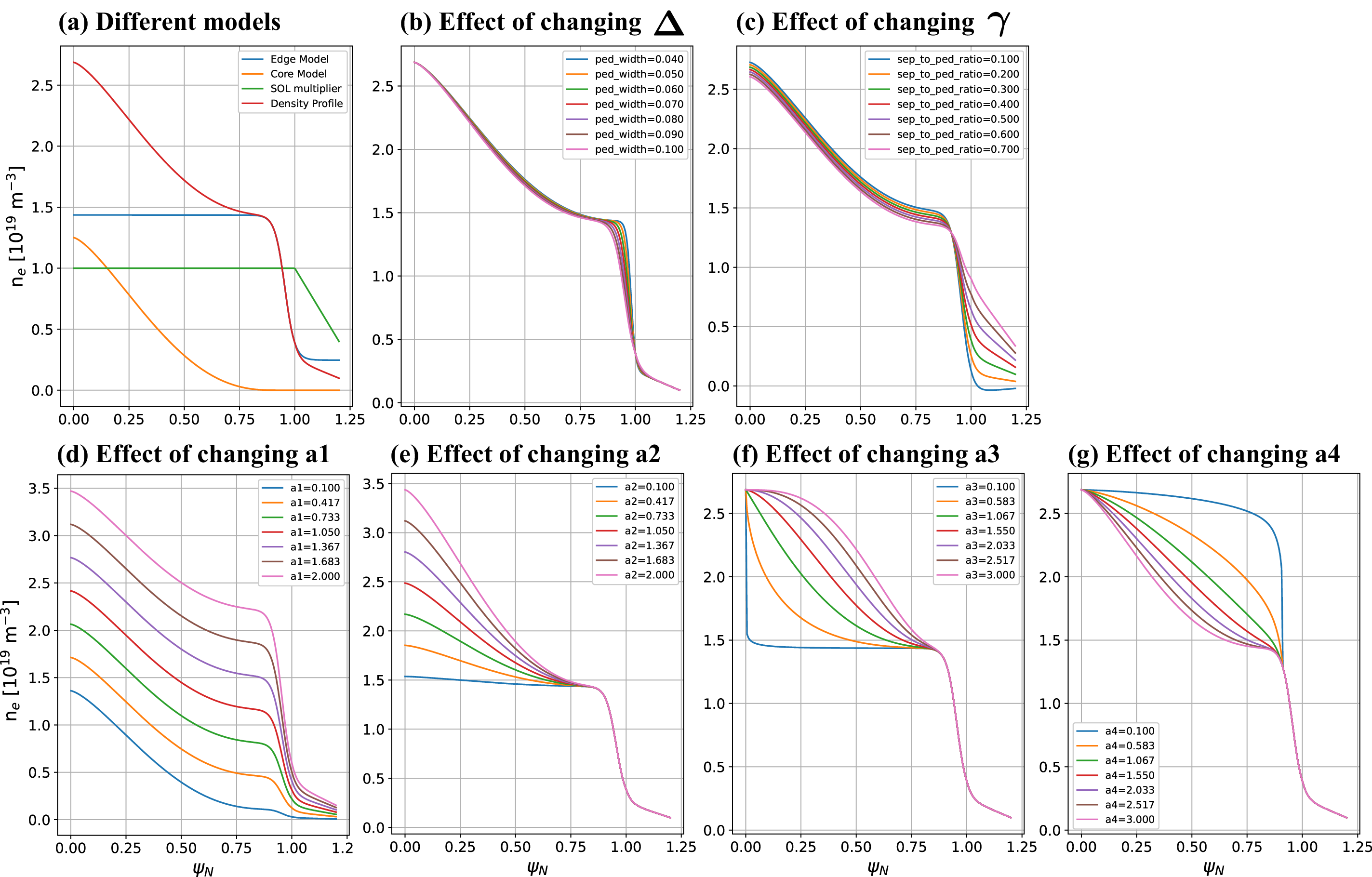}
\caption[figure visual]{The figure shows how sensitive the model is to the fitting parameters. $\boldsymbol{(a)}$ Visualizes the model comprising the fitting function for KSTAR $\#$33573 t=\SI{2.679}{\second}, where $\Delta=0.09$, $\gamma=0.3$, $a1= 1.29$, $a2= 1.25$, $a3= 1.39$, and $a4= 2.60$. $\boldsymbol{(b)-(g)}$ Show the sensitivity of the fitting model to the six fitting parameters.}
\label{fig:profile_visual}
\end{center}
\end{figure*}

\noindent The effect of the six fitting parameters on the profile shape is illustrated in Fig. \ref{fig:profile_visual}. Although it makes the fitting function less expressive, we fix $\gamma = 3.0 \times 10^{-1}$ and $\Delta = 9.0 \times 10^{-2}$, as we only have the five measurements from two-colored interferometer (TCI), which measures the line-averaged density, to conduct the fitting procedure.

\subsection{Profile fitting with EFIT and TCI}
\label{subsec:TCI}

\begin{figure}[htbp]
\begin{center}
\includegraphics[width=\linewidth]{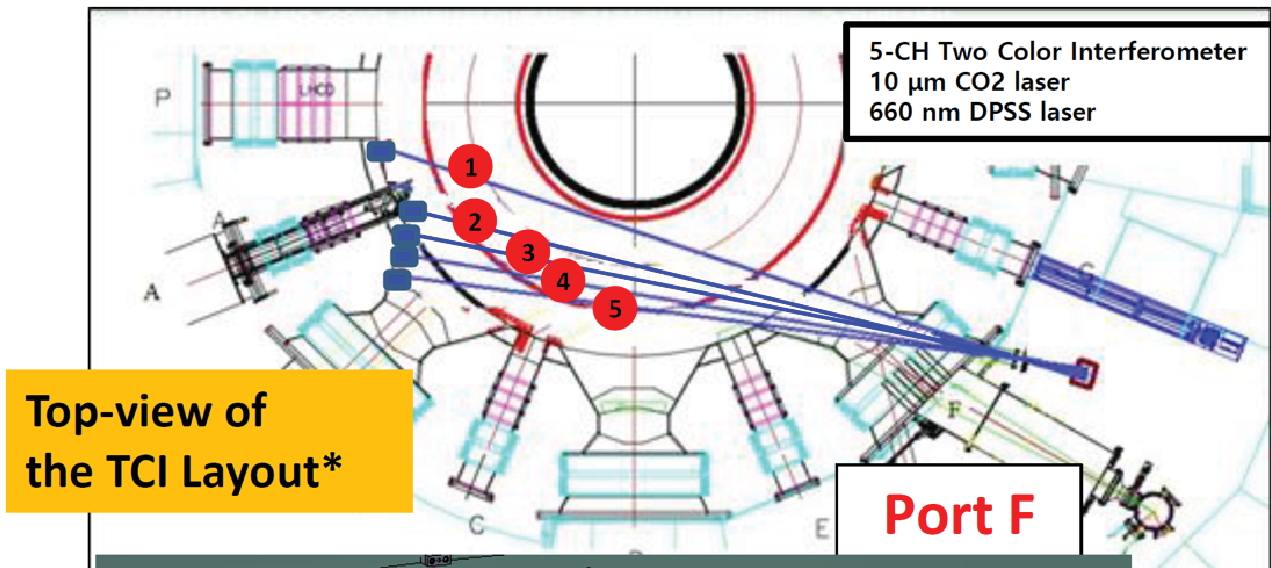}
\caption{Illustration of the KSTAR TCI line of sight. The line of sight resides on the Z=0 plane, and there are five channels in total. The figure is from an internal report at the Korea Institute of Fusion Energy (KFE). Details of the KSTAR TCI system are described in \cite{paper:juhn}.}
\label{fig:los}
\end{center}
\end{figure}

\begin{figure}[htbp]
\begin{center}
\includegraphics[width=\linewidth]{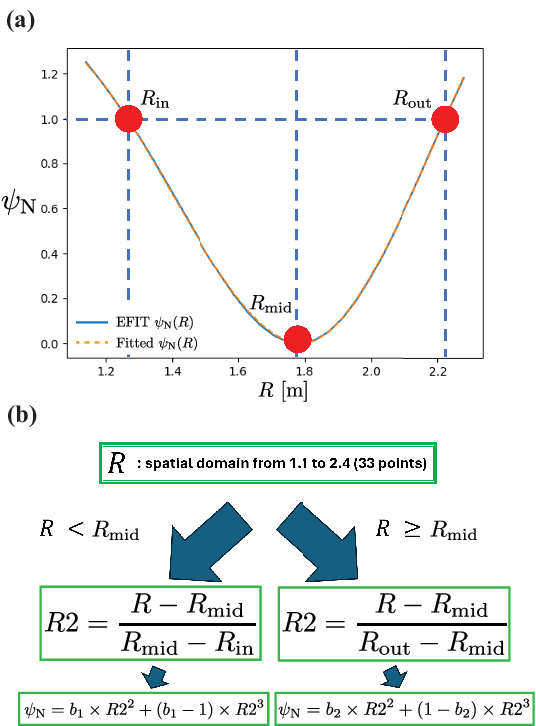}
\caption{$\boldsymbol{(a)}$ Illustration of the normalized poloidal flux function, $\psi_\textrm{N}$, mapped to the real-space domain on the $Z = 0$ plane. The radial position of the minimum $\psi_\textrm{N}$ at $Z = 0$ is defined as $R_{\mathrm{mid}}$ (in meters). The inboard and outboard radial positions of the last closed flux surface (LCFS) at $Z = 0$ are denoted by $R_{\mathrm{in}}$ and $R_{\mathrm{out}}$, respectively (in meters). $\boldsymbol{(b)}$ Description of the $\psi_\textrm{N}$ mapping procedure. The radial grid is normalized such that $R_{\mathrm{in}}$ and $R_{\mathrm{out}}$ correspond to -1 and 1, respectively, to facilitate the determination of fitting coefficients $b_1$ and $b_2$. These coefficients are obtained by minimizing the fitting error between $\psi_\textrm{N}$ and the model function. Once fitted, the normalized grid is mapped back to the real-space radial coordinate $R$.}
\label{fig:EFIT mapping}
\end{center}
\end{figure}

The KSTAR TCI system measures the line-averaged density along the line of sight on the $Z=0$ plane of the device as illustrated in Fig. \ref{fig:los}. Inversely, we can also estimate the line-averaged density from the density profile reconstructed by the fitting function described in Sec. \ref{subsec:model} on the $\psi_{N}$ coordinate. The fitting procedure is to find the coefficients $a_1$, $a_2$, $a_3$, and $a_4$, minimizing the error between the line-averaged density measured by TCI and estimated from the density profile. Since the discrete integration along the line of sight should be conducted on the real axis whose dimensionality is in meters, we need a mapping of $\psi_\textrm{N}$ to the real space. An example of the mapping is illustrated in Fig. \ref{fig:EFIT mapping}. (a). The mapping can be parametrized by the two coefficients $b_1$ and $b_2$ as described in Fig. \ref{fig:EFIT mapping}. (b). In summary, the $n_e$ procedure consists of 

\begin{enumerate}
    \item Parameterize the mapping from $\psi_\textrm{N}$ to real space.
    \item Perform discrete integration of the $n_e$ profile along the TCI line of sight.
    \item Identify the fitting coefficients $a_1$ through $a_4$ that minimize the discrepancy between the measured and reconstructed line-averaged densities.
\end{enumerate}

The parametrization of the flux function $\psi_\textrm{N}$ takes only two parameters, $b_1$ and $b_2$, and can be done within the cycle time of KSTAR PCS, which is on the order of $\mu s$. In contrast, the computational time of the fitting of the $n_\textrm{e}$ profile takes time on the order of ms, which is not feasible for real-time purposes. To let our controller exploit the speed of PCS, we decided to accelerate the fitting procedure by using a neural network. By doing so, we can make a pedestal-top $n_\trm{e}$ controller to respond to the target in real-time.

\subsection{Neural network acceleration for control}
\label{subsec:NN}

There are ten inputs of the neural network, ($b_1$, $b_2$, $R_{\trm{mid}}$, $R_{\trm{in}}$, $R_{\trm{out}}$, $\overline{n}_{\trm{e1}}$, $\overline{n}_{\trm{e2}}$, $\overline{n}_{\trm{e3}}$, $\overline{n}_{\trm{e4}}$, $\overline{n}_{\trm{e5}}$). $b_1$ and $b_2$ are the fitting parameters to map normalized poloidal flux ($\psi_{\mathrm{N}})$ to the real-space, $R_{\trm{mid}}$ is the radial location of minimum $\psi_{\mathrm{N}}$ on the $Z=0$ plane (in meters), $R_{\trm{in}}$ and $R_{\trm{out}}$ are the radial locations of the last closed flux surface (LCFS) on in- and outboard (in meters), and $\overline{n}_{\trm{e1}\sim5}$ are the line-averaged density measured by TCI (\SI{1e19}{\per\cubic\meter}). $R_{\trm{mid}}$, $R_{\trm{in}}$, and $R_{\trm{out}}$ are reconstructed by real-time EFIT (rtEFIT). The neural network outputs are ($a_1$, $a_2$, $a_3$, $a_4$), which are the fitting coefficients of the $n_e$ profile fitting function. Hence, the neural network is basically replacing the $n_\trm{e}$ profile fitting process. It has two hidden layers, each with 20 nodes, with batch normalization and \textit{tanh} activation function. \textit{ReLU} activation function is used for the output layer with batch normalization. The training is done to minimize the mean squared error (MSE) with the Adam optimizer. The total number of parameters of the neural network is 900, the minimized MSE of the validation set is \SI{4.05e-2}{}, and the minimized MSE of the training set is \SI{3.42e-2}{}. The number of hidden layers, the number of nodes in each hidden layer, and the activation functions are selected to maximize the $R^2$ values of $a_1$ and $a_2$ of the validation datasets. The reason why we are trying to maximize the $R^2$ values of $a_1$ and $a_2$ of the validation datasets is that the control target is pedestal-top $n_\trm{e}$, which is represented by $a_1$, and we have a plan to control core $n_\trm{e}$ as well which is relevant to $a_2$ representing the difference between core and pedestal-top $n_e$. Since $a_3$ and $a_4$ only determine the shape of the profile in the core region, we put less importance on these parameters.

\begin{figure*}[htbp]
\begin{center}
\includegraphics[width=\linewidth]{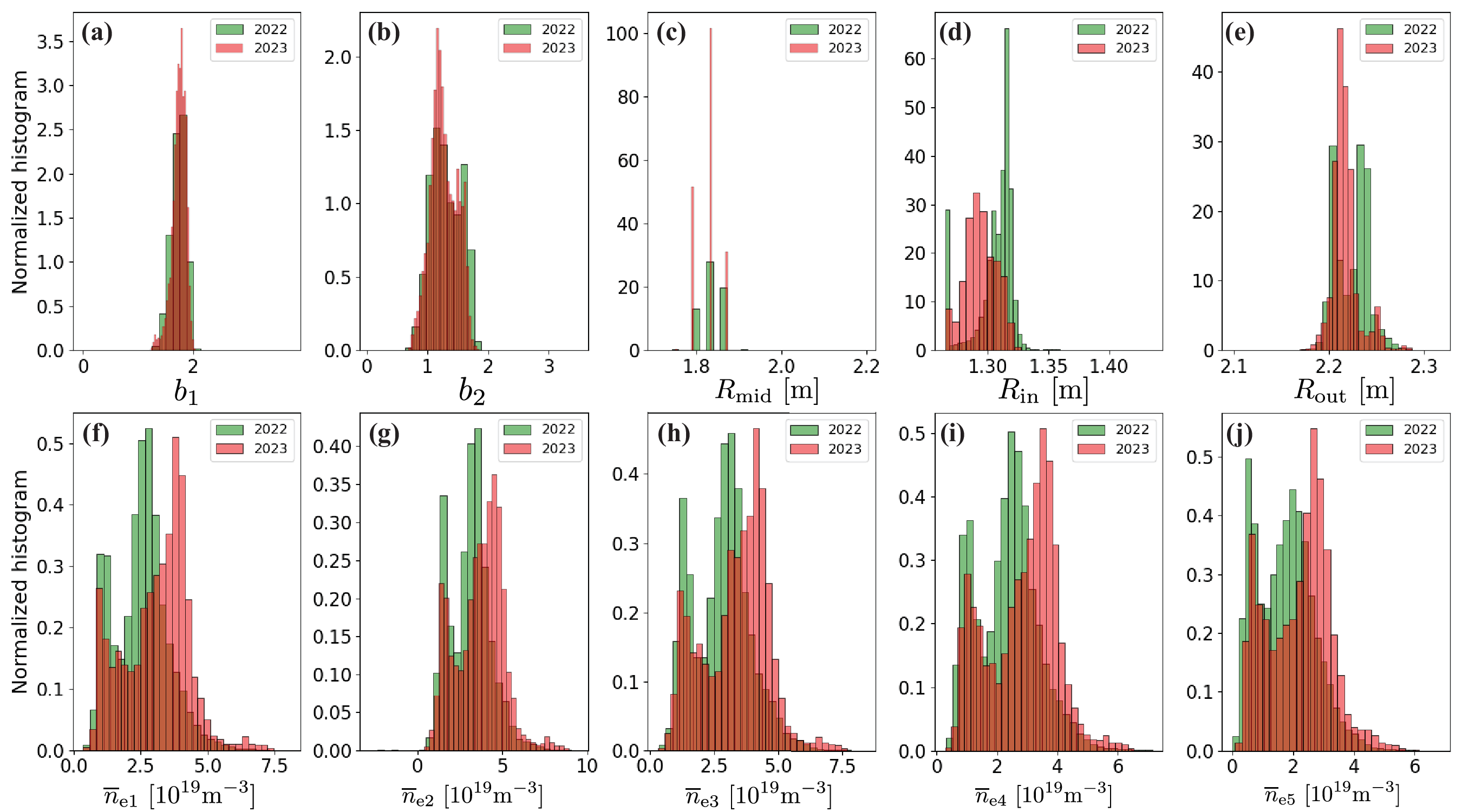}
\caption{Histograms of the 2022 and 2023 KSTAR datasets. $\boldsymbol{(a)-(b)}$ Fitting coefficients of $\psi_N$ at $Z = 0$, denoted as $b_1$ and $b_2$, respectively. $\boldsymbol{(c)}$ Radial position of the minimum $\psi_{\mathrm{N}}$ at Z=0 plane, $R_{\mathrm{mid}}$ (in meters), reconstructed by rtEFIT. $\boldsymbol{(d)}$ Inboard radial position of the last closed flux surface (LCFS), $R_{\mathrm{in}}$ (in meters), also from rtEFIT. $\boldsymbol{(e)}$ Outboard LCFS radial position, $R_{\mathrm{out}}$ (in meters), from rtEFIT. $\boldsymbol{(f)-(j)}$ Line-averaged electron densities, $\overline{n}_{\mathrm{e1}}$ to $\overline{n}_{\mathrm{e5}}$, measured by the TCI diagnostic. A notable distributional shift is observed in 2023 due to replacing the lower divertor with tungsten, highlighting the challenge and necessity of ensuring that the machine learning predictor (MLP) generalizes well across experimental upgrades.}
\label{fig:hist}
\end{center}
\end{figure*}

\begin{figure*}[htbp]
\begin{center}
\includegraphics[width=\linewidth]{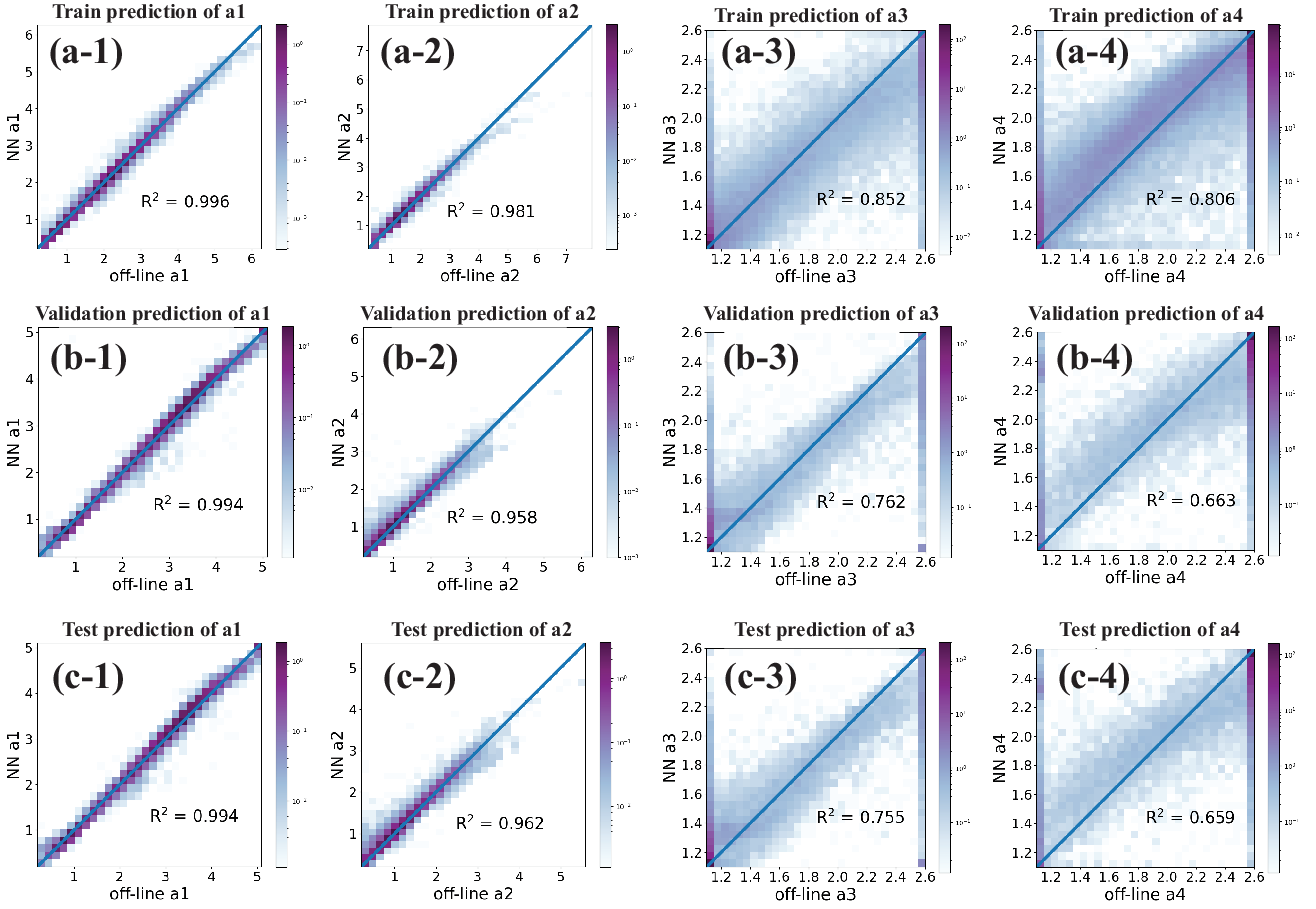}
\caption{The comparison of the four fitting parameters $a1$ to $a4$ between reconstructed offline and by neural network for $\boldsymbol{(a)}$ the training dataset from the 2022 KSTAR campaign, $\boldsymbol{(b)}$ validation dataset from the 2023 KSTAR campaign, and $\boldsymbol{(c)}$ test dataset from the 2023 KSTAR campaign. Since validation and test datasets are randomly split from the 2023 KSTAR campaign data, they have almost the same $R^2$ values. Training results have slightly higher $R^2$ values for all four fitting parameters.}
\label{fig:r_square}
\end{center}
\end{figure*}

\begin{figure}[htbp]
\begin{center}
\includegraphics[width=\linewidth]{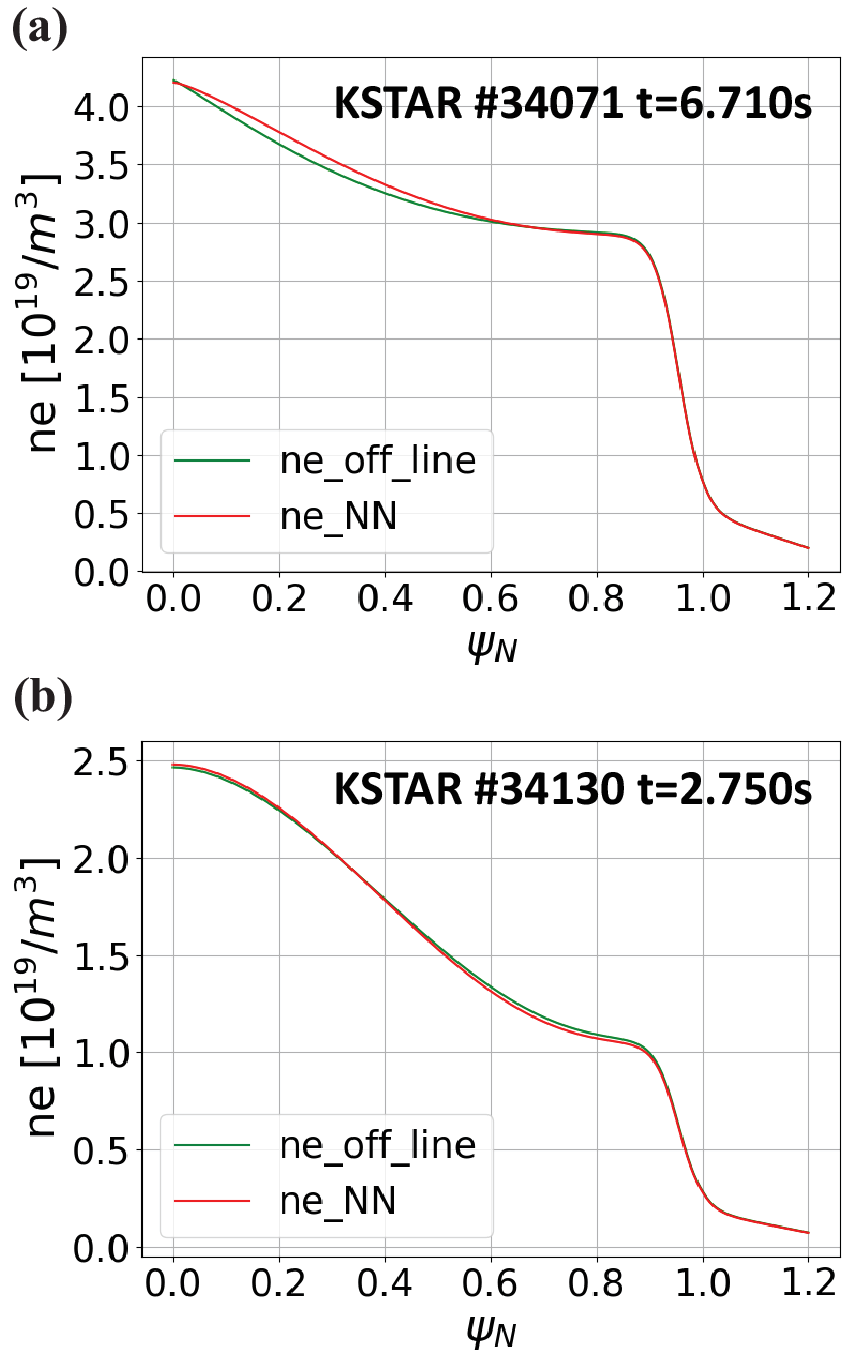}
\caption{Comparison of the reconstructed profiles through the offline method and neural network for $\boldsymbol{(a)}$ KSTAR $\#$34071 t=\SI{6.710}{\second} and $\boldsymbol{(b)}$ KSTAR $\#$34130 t=\SI{2.750}{\second}. The red plot represents the neural network predicted one, and the green one is reconstructed offline. The spatial domain is the normalized poloidal flux function.}
\label{fig:NN_results}
\end{center}
\end{figure}

Training, validation, and test datasets are respectively comprised of 1,406 shots with 93,812 time-slices from the 2022 KSTAR campaign, 375 shots with 33,591 time-slices from the 2023 KSTAR campaign, and 376 shots with 33,592 time-slices from the 2023 KSTAR campaign. The data from the flat-top region is selected, and diagnostics failure cases are pruned. The histogram of the dataset is in Fig. \ref{fig:hist}. The validation and test datasets are comprised of 2023 KSTAR data, which is in a different campaign year than the training data, to see how much our model can generalize the results to changes in the experimental setup, such as the upgrade of lower divertor tiles to be Tungsten between the 2022 and 2023 campaigns. The radial position of in- and outboard LCFS on the $Z=0$ plane and the measured line-averaged densities show a notable difference between the two years, implying that changing the lower divertor tiles to Tungsten impacted the operational scenarios to have different plasma shapes and electron density levels. The $R^2$ training, validation, and test dataset values are illustrated in Fig. \ref{fig:r_square}. The median and mean absolute percentage errors of $n_\mathrm{e}$ on the test dataset are \SI{1.85}{\percent} and \SI{2.38}{\percent}, respectively. The offline and neural network reconstructed $n_\mathrm{e}$ profiles are compared in Fig. \ref{fig:NN_results}. The computational time in the KSTAR PCS of the whole fitting process with the neural network is about 120 $\mu s$, less than the CPU's cycle time, 500 $\mu s$. We conclude that the accuracy and speed of the algorithm satisfy our requirements for the control experiments.

\section{Pedestal-top electron density control}
\label{sec:necontrol}

\begin{figure}[htbp]
\begin{center}
\includegraphics[width=\linewidth]{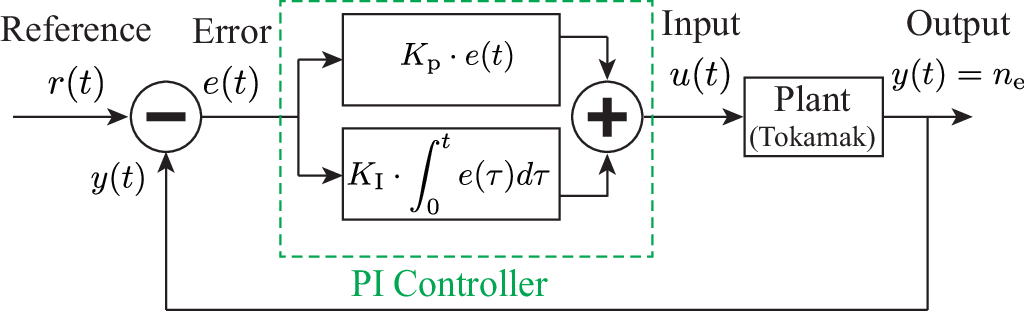}
\caption{Schematic view of the closed-loop system. The output $y(t)$ is the electron density $n_{\mathrm{e}}$ at $\psi_{\mathrm{N}} = 0.89$, and the input $u(t)$ is either the IVCC currents or the voltage applied to the main gas puff. The input $u(t)$ is computed by the PI controller based on the error $e(t) = r(t) - y(t)$.}
\label{fig:cl_loop}
\end{center}
\end{figure}

\subsection{System identification to design the PI controller}
\label{subsec:SI}

\begin{figure}[htbp]
\begin{center}
\includegraphics[width=0.6\linewidth]{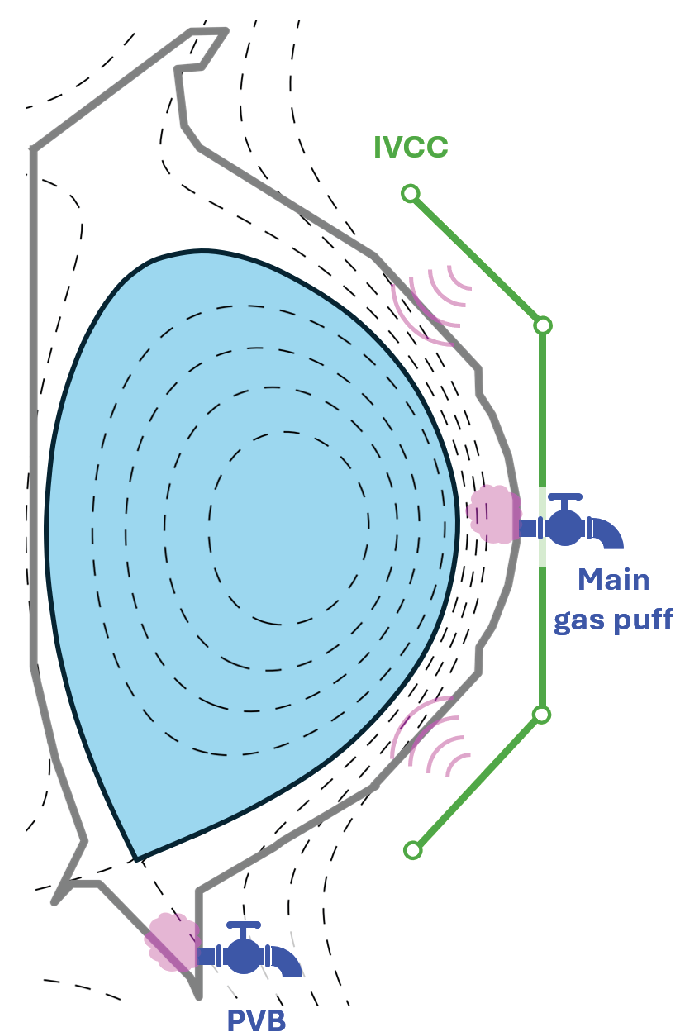}
\caption{Image courtesy: CheolSik Byun. Cross-section of KSTAR showing the location of actuators. There are 12 sets of IVCC, four at the top, four in the middle, and four at the bottom. The coil sets can make the RMP with different mode numbers and phasing. The main gas puff injects gas in the midplane, and the PVB injects gas near the divertor. We approximated the plasma response to the main gas puff with the plasma response to the PVB with $D_2$ fuel because of the limited experimental run-time and lack of plasma response for the main gas puff in the reference discharge.}
\label{fig:crosssection}
\end{center}
\end{figure}

\begin{figure}[htbp]
\begin{center}
\includegraphics[width=\linewidth]{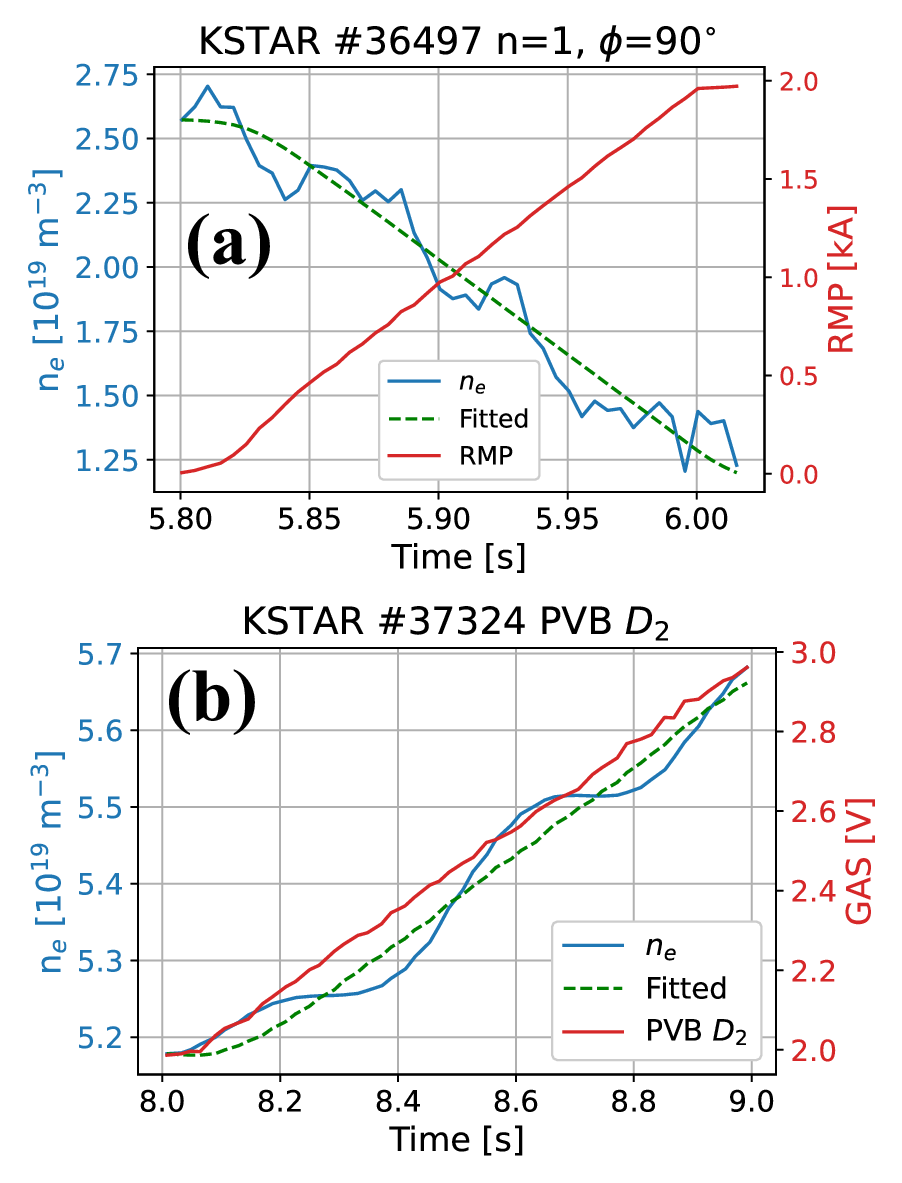}
\caption{System identification results for $\boldsymbol{(a)}$ n=1, $\phi_{\trm{TM}}=\phi_{\trm{MB}}=90^{\circ}$ RMP and $\boldsymbol{(b)}$ PVB with $D_2$ fuel. The actuators are plotted in red lines. The blue lines are the $y(t) = n_\trm{e}(t)$ at $\psi_\trm{N}=0.89$. Green dotted lines are the fitted $n_\trm{e}(t)$ adopting the first-order model in Eq. \ref{eq:impulse}, which minimized the error between $h(t) * u(t)$ and $y(t)$. For the PVB $D_2$ case, a Butterworth filter with a cutoff frequency of 3 Hz was used for the $n_\trm{e}$ at $\psi_\trm{N}=0.89$.}
\label{fig:SI}
\end{center}
\end{figure}

Terminologies in control theory are described in the \ref{apx:sys_id}. For our problem, the input $u(t)$ is the current of IVCC, which applies RMP to the plasmas, and the voltage applied to the main gas puff, which is the main actuator of KSTAR gas control. The output $y(t)$ is the $n_\trm{e}$ at $\psi_\trm{N} = 0.89$ reconstructed by the method described in Sec. \ref{sec:nerecon}. With the impulse response in Eq. \ref{eq:impulse} obtained by our first-order model in Eq. \ref{eq:dynamics}, we could conduct the discrete convolution between the input $u(t)$ and the impulse response $h(t)$. $K$ and $T$ can be estimated to minimize the error between the output $y(t)$ from the experiment and the one we calculated from the convolution. We dedicate an experimental shot to conduct system identification for n=1, $\phi_{\trm{TM}}=\phi_{\trm{MB}}=90^{\circ}$ RMP, but approximated the dynamics of main gas puff with the piezo-electric valve bottom (PVB) with $D_2$ fuel of the reference discharge because of the limited experimental run-time and lack of plasma response with respect to main gas puff in the reference discharge. Uncalibrated TCI data were used to determine $K$ and $T$ for RMP and to design the controller, despite the control experiments and system identification for the PVB being conducted with calibrated data. The $K$ and $T$ for RMP from calibrated and uncalibrated TCI are in the Table. \ref{tb:RMP}, and they are still in a similar order. We designed the controller with the system identification results for RMP and PVB and wanted to see if the controller gains are generalizable for other actuator configurations, such as RMP with different mode numbers and phasing, or different gas actuator. The system identification results are illustrated in Fig. \ref{fig:SI}. The illustration of the actuators inside of KSTAR is in the Figure. \ref{fig:crosssection}.

\begin{table}[htbp]
  \centering
  \begin{minipage}{0.48\textwidth}  

    \footnotesize                           
    \setlength{\tabcolsep}{2pt}             
    \renewcommand{\arraystretch}{0.9}       

    \caption{System identification results of
             $n{=}1$, $\phi_{\textrm{TM}}=\phi_{\textrm{MB}}=90^\circ$
             RMP with calibrated and uncalibrated TCI, and of PVB with $D_2$ fuel.}
    \label{tb:RMP}

    \begin{tabularx}{\linewidth}{@{}
        >{\centering\arraybackslash}m{11mm}   
        >{\centering\arraybackslash}X         
        >{\centering\arraybackslash}c
        >{\centering\arraybackslash}c
        @{}}
      \toprule
      & TCI & $K$ & $T$ \\
      \midrule
      \multirow{2}{*}{\centering\textbf{RMP}} &
        Uncalibrated & \num{-7.44e-1} & \num{2.34e-2} \\
      & Calibrated & \num{-1.25}    & \num{1.00e-2} \\[1pt]
      \textbf{PVB} &
        Calibrated & \num{4.45e-1}  & \num{9.09e-2} \\
      \bottomrule
    \end{tabularx}
  \end{minipage}
\end{table}

\subsection{PI controller design with pole placement}
\label{subsec:PI}

The closed-loop can change the system's dynamics such that the system's output can follow the reference value $r(t)$ while it satisfies the control requirements by properly designing the controller. The error $e(t) = r(t) - y(t)$ is fed into the controller, and it gives input to the plant $u(t)$. The PI controller was used for our experiments, which calculates 

\begin{align}
u(t) = K_\trm{p} \cdot e(t) + K_\trm{I} \cdot \int^{t}_{0} e(\tau) d\tau.
\label{eq:input_expression}
\end{align}

The integral term gives us the error history so that we can have the input $u(t)$ even if we reach the target, and let us have zero steady-state error for a step target. If we do not have this term, $u(t)$ would become zero instantly when the $y(t)$ reaches the target $r(t)$. Even if the integral term lets us reach the step target if enough time has passed, the proportional term is necessary to reach a dynamical target faster. The derivative term can help reduce overshoot in the output $y(t)$, but it is sensitive to noise, making it difficult to tune. For this reason, it is excluded. Since the P, PI, or PID controllers are simple but effective to follow the target for various cases, it has been widely used in fusion research \cite{paper:kolemen2010strike, paper:park2013investigation, thesis:JWJuhn2013study, paper:fahmy2016multivariable, paper:ravensbergen2017density, paper:blanken2019model}. The schematic view of the closed-loop system is shown in Figure. \ref{fig:cl_loop}.

$C(s)$ is the transfer function of the controller, $H(s)$ is the transfer function of the plant expressed by Eq. \ref{eq:transfer_laplace}, $R(s)$ is the laplace transformed $r(t)$, and $G(s)$ is the transfer function of the closed-loop, where

\begin{align}
C(s) &= K_\trm{p} + K_\trm{I} \frac{1}{s},  \label{eq:controller}
\\ Y(s) &= \frac{HC}{1+HC} \cdot R(s)  \nonumber
\\ & \equiv G(s) \cdot R(s), \label{eq:cl_transfer}
\\ G(s) &= \frac{K/T(K_\trm{p}\cdot s + K_\trm{I})}{s^2 + [(K \cdot K_\trm{p}+1)/T]s + K \cdot K_\trm{I}/T}. \label{eq:gs}
\end{align}

\begin{figure}[htbp]
\begin{center}
\includegraphics[width=\linewidth]{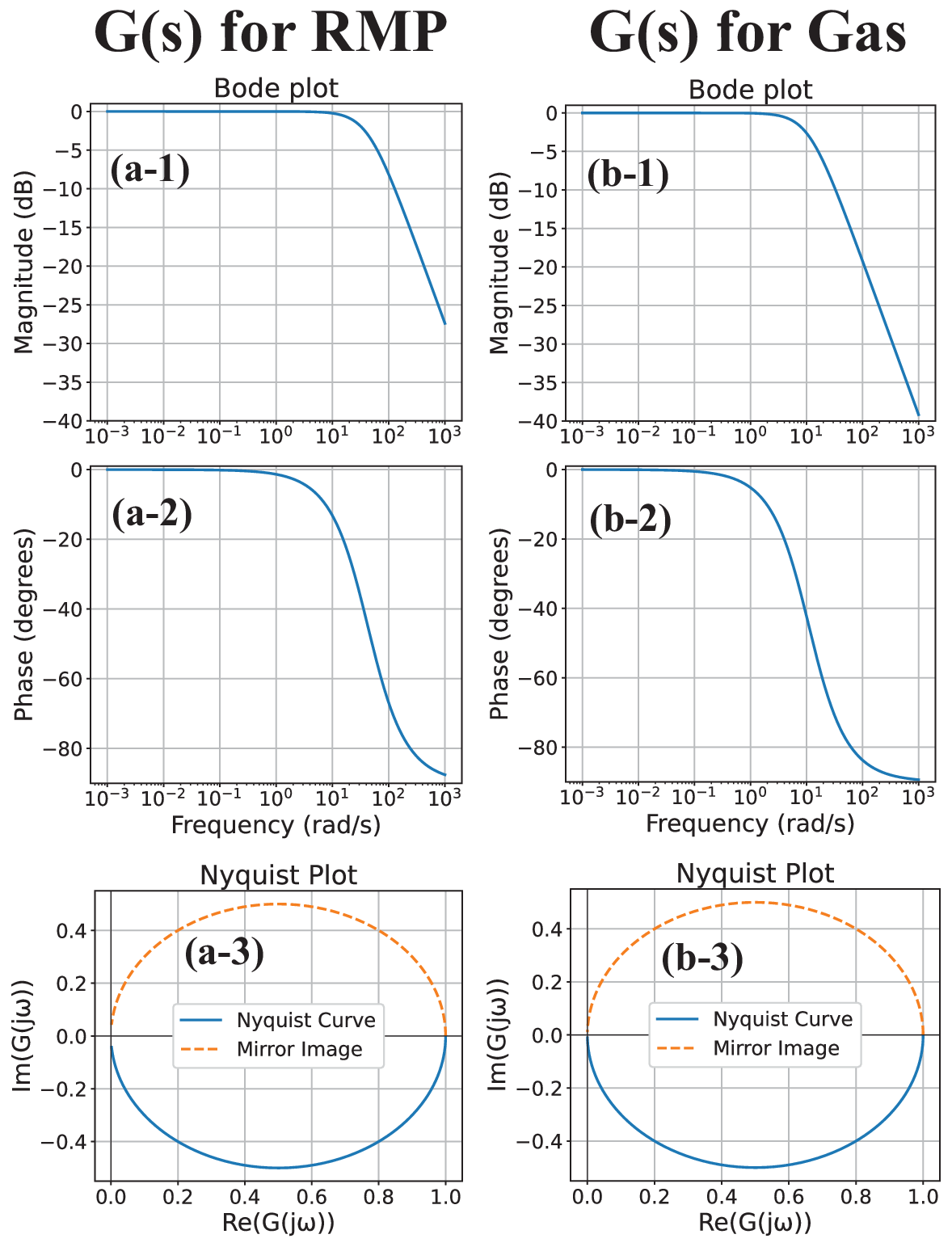}
\caption{Bode and Nyquist plots of the designed closed-loop transfer function of $\boldsymbol{(a)}$ n=1, $\phi_{TM} = \phi_{MB} = 90^{\circ}$ RMP and $\boldsymbol{(b)}$ PVB with $D_2$ fuel.}
\label{fig:bode}
\end{center}
\end{figure}

Placing the pole of the closed-loop transfer function $G(s)$ in the same place as the plant's pole can result in a robust controller when the plant has fast poles \cite{book:Rowley}. For our case, the pole of the plant is $-1/T \sim -10^2$, which can be treated as a fast pole compared to the bandwidth we want the controller to have ($\sim$10 rad/s). Since our first-order model in Eq. \ref{eq:dynamics} has simplified the dynamics of the plasmas, we did not specify the control requirements of settling time, rising time, and overshoot, but applied a simple rule to place poles. By using the pole placement method, our controller has the gain of $K_\trm{p} = \frac{1}{K}$ and $K_\trm{I} = \frac{1}{KT}$. The controller gain sets obtained by the pole placement are $K_\trm{p} = -1.35$ and $K_\trm{I} = -5.74 \times 10$ for RMP and $K_\trm{p} = 2.25$ and $K_\trm{I} = 2.47 \times 10$ for main gas puff, which are calculated by the $K$ and $T$ in Figure. \ref{fig:SI}. It shows that the fitted line matches well with the $n_\mathrm{e}$ reconstructed during the experiments. Bode and Nyquist plots of the closed-loop transfer function $G(s)$ in Fig. \ref{fig:bode} let us know the bandwidth and stability of our controller. The controller using RMP as an actuator has a larger bandwidth than the one with main gas puff. This implies that the first one would try to reach the target more aggressively but have a higher chance of overshooting. In addition, the designed controllers have almost \SI{180}{\degree} phase margin and infinite gain margin.

\subsection{Pedestal-top electron density control result using a single actuator, RMP}
\label{subsec:results_RMP}

\begin{figure}[htbp]
\begin{center}
\includegraphics[width=\linewidth]{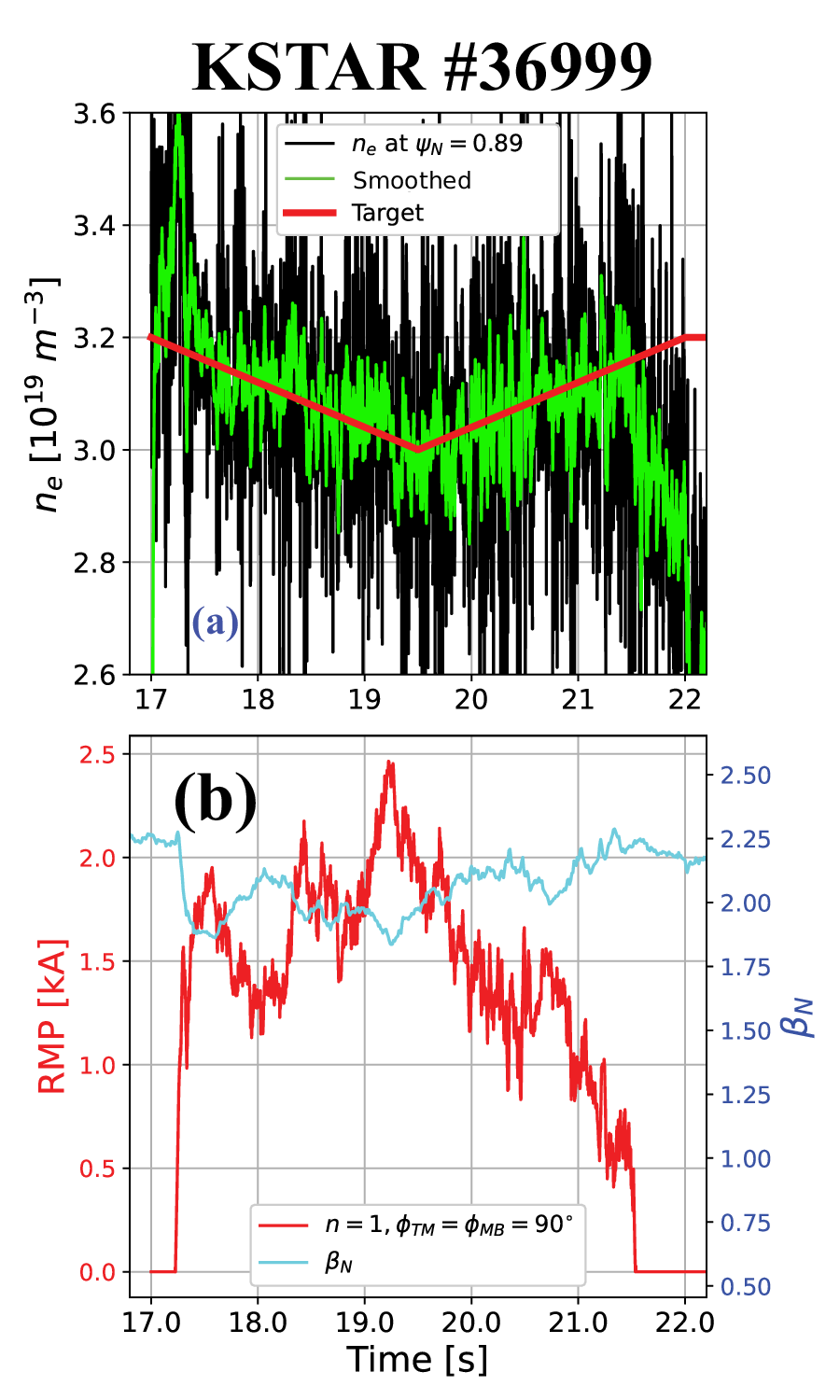}
\caption{Experimental result of controlling $n_\trm{e}$ at $\psi_\trm{N}=0.89$ using n=1, $\phi_{\trm{TM}}=\phi_{\trm{MB}}=90^{\circ}$ RMP as an actuator for KSTAR $\#$36999. $\boldsymbol{(a)}$ The black line gives $n_\trm{e}$ at $\psi_\trm{N}=0.89$ reconstructed by the neural network, and the lime line is the smoothed value of $n_\trm{e}$. The red line gives the control target. The shot length was 22 s. The smoothing is done by using Eq. \ref{eq:LPF}. $\boldsymbol{(b)}$ The red line shows RMP current, and the cyan line shows $\beta_N$ calculated by EFIT04 (with linear drift corrected magnetics). The controller was turned on at \SI{17}{\second} and turned off at \SI{22}{\second}. As shown in the figures, the controller reacts to minimize the difference between the target and the $n_\trm{e}$, and the $\beta_\trm{N}$ value drops when we apply the RMP.}
\label{fig:shin}
\end{center}
\end{figure}

For the first trial, pedestal-top $n_\trm{e}$ was controlled only by using n=1, $\phi_{\trm{TM}}=\phi_{\trm{MB}}=90^{\circ}$ RMP, and one of the control results is plotted in Figure. \ref{fig:shin}. The control experiment was conducted from \SI{17}{\second} to \SI{22}{\second} of KSTAR $\#$36999 with $I_\trm{P}=0.5$MA and $B_\trm{T}=1.9$T. The $n_\trm{e}$ at $\psi_\trm{N}=0.89$ reconstructed by the neural network has been smoothed with the formula,

\begin{align}
&n_\trm{{e, smooth}}(t) \nonumber 
\\= &n_\trm{{e, smooth}}(t-1)+\frac{\Delta t}{\Delta t + \tau} \cdot [ n_{\trm{e}}(t) - n_\trm{{e, smooth}}(t-1) ], \label{eq:LPF}
\end{align}

\noindent where $\Delta t$ is the sampling period, which is \SI{500}{\micro\second} and $\tau$ is set to be \SI{10}{\milli\second}. $K_\trm{p}$ and $K_\trm{I}$ for RMP are set to the values calculated in Sec. \ref{subsec:PI}. The control target linearly decreases, then increases. The density followed the target even if there was some overshoot. The overshoot may come from the fact that the system identification was done using the uncalibrated TCI data. If we compare the gain sets from calibrated and uncalibrated ones, the uncalibrated one gives us larger $K_\trm{p}$ and smaller $K_\trm{I}$, which would result in larger overshoot. The observed drop in $\beta_{\mathrm{N}}$, coinciding with the increase in RMP currents, indicates that the density reduction is caused by the RMP. RMP currents are zero at the beginning because the density is initialized to zero at the algorithm's beginning. The absolute percentage errors between the $n_\trm{e}$ at $\psi_\trm{N}=0.89$ and the target have a \SI{1.72}{\percent} median and a \SI{2.72}{\percent} average value in KSTAR $\#$36999 during when the controller was on.

\subsection{Pedestal-top electron density control result using multiple actuators, RMP, and main gas puff}
\label{subsec:results_combo}

\begin{figure}[htbp]
\begin{center}
\includegraphics[width=0.8\linewidth]{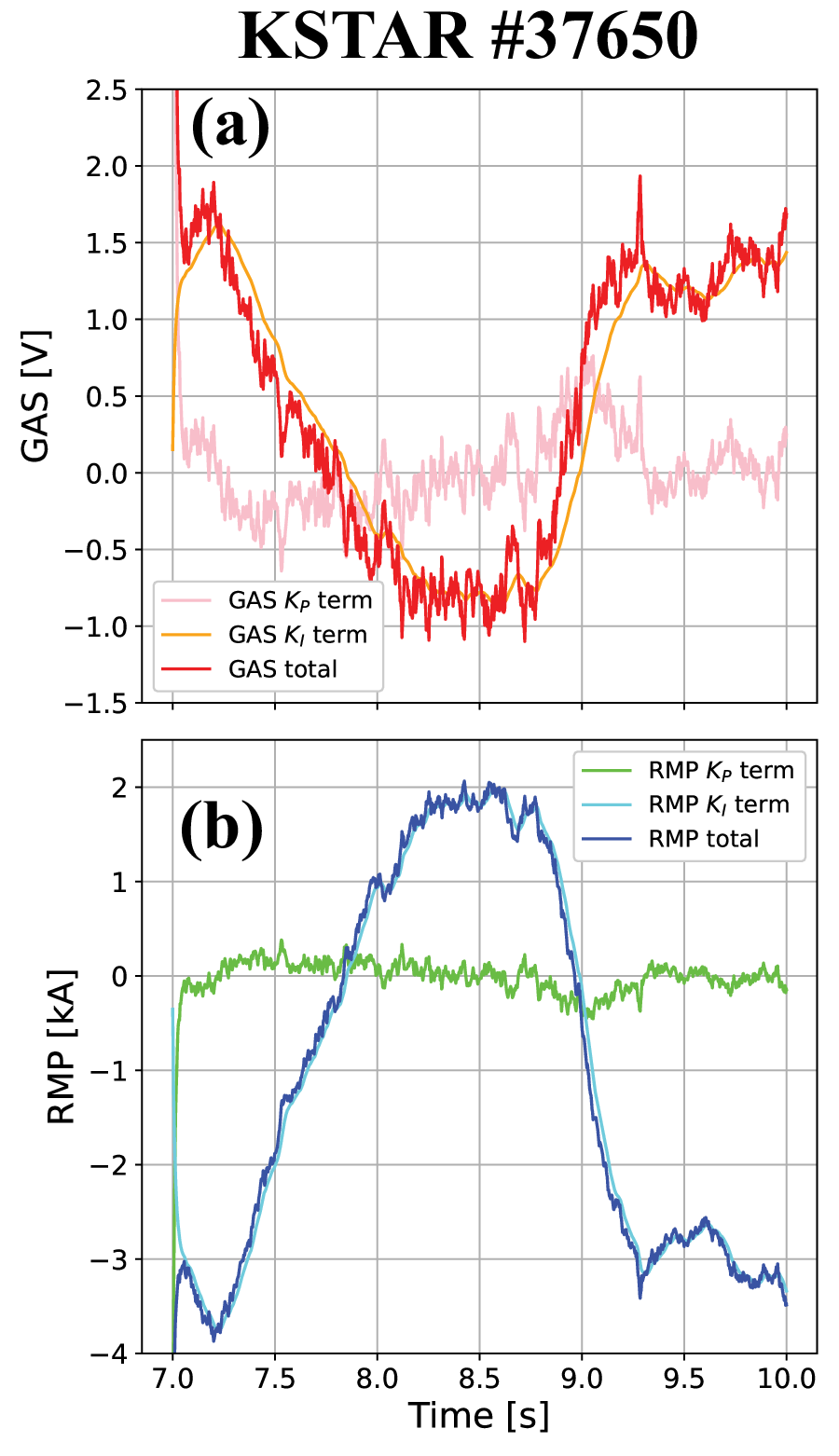}
\caption{$\boldsymbol{(a)}$ The pink line shows the $K_\trm{p} \cdot e(t)$, and the orange line is the $K_\trm{I} \cdot \int^{t}_{0}e(\tau)d\tau$ for the main gas puff. The red line represents the sum of them calculated by the PI controller for the main gas puff, where $K_\trm{p}$ and $K_\trm{I}$ are decided by the system identification with PVB with $D_2$ fuel. $\boldsymbol{(b)}$ The green line shows the $K_\trm{p} \cdot e(t)$, and the cyan line is the $K_\trm{I} \cdot \int^{t}_{0}e(\tau)d\tau$ for n=2, $\phi_{\trm{TM}}=\phi_{\trm{MB}}=90^{\circ}$ RMP. The blue line represents the sum of them calculated by the PI controller for n=2, $\phi_{\trm{TM}}=\phi_{\trm{MB}}=90^{\circ}$ RMP where $K_\trm{p}$ and $K_\trm{I}$ are decided by the system identification with n=1, $\phi_{\trm{TM}}=\phi_{\trm{MB}}=90^{\circ}$ RMP. $K_\trm{p}$ and $K_\trm{I}$ for both cases are in the Section. \ref{subsec:PI}.}
\label{fig:err}
\end{center}
\end{figure}

After we got the promising results of the pedestal-top $n_\trm{e}$ control only using $n=1, \ \phi_{\trm{TM}}=\phi_{\trm{MB}}=90^{\circ}$ RMP, we decided to add main gas puff as an actuator as well to have a more complete controller. Since RMP only decreases the $n_\trm{e}$, using only RMP as an actuator can not let the $n_\trm{e}$ reach the target if the target is higher than the $n_\trm{e}$ from the beginning of the control. The same thing would happen with gas as well if the target is lower than the $n_\trm{e}$, as it can only increase the $n_\trm{e}$. The implementation of the integrated controller was achieved by using the fact that the signs of the two inputs $u_{\trm{RMP}}(t)$ and $u_{\trm{gas}}(t)$ are different from each other. This is because the $u(t)$ is dominantly determined by the integration term since $K_\trm{I}$ is larger than $K_\trm{p}$ for both actuators, and the flipped sign of the gains of the two actuators lets them have mutually exclusive $u(t)$ values. This is illustrated in Figure. \ref{fig:err}. It means that if $u_{\trm{RMP}}(t) > 0$ then $u_{\trm{gas}}(t) < 0$ and vice versa. By clamping the inputs to be larger than zero, we implemented the integrated controller exclusively using one of the two actuators, the main gas puff and RMP.

\begin{figure}[htbp]
\begin{center}
\includegraphics[width=\linewidth]{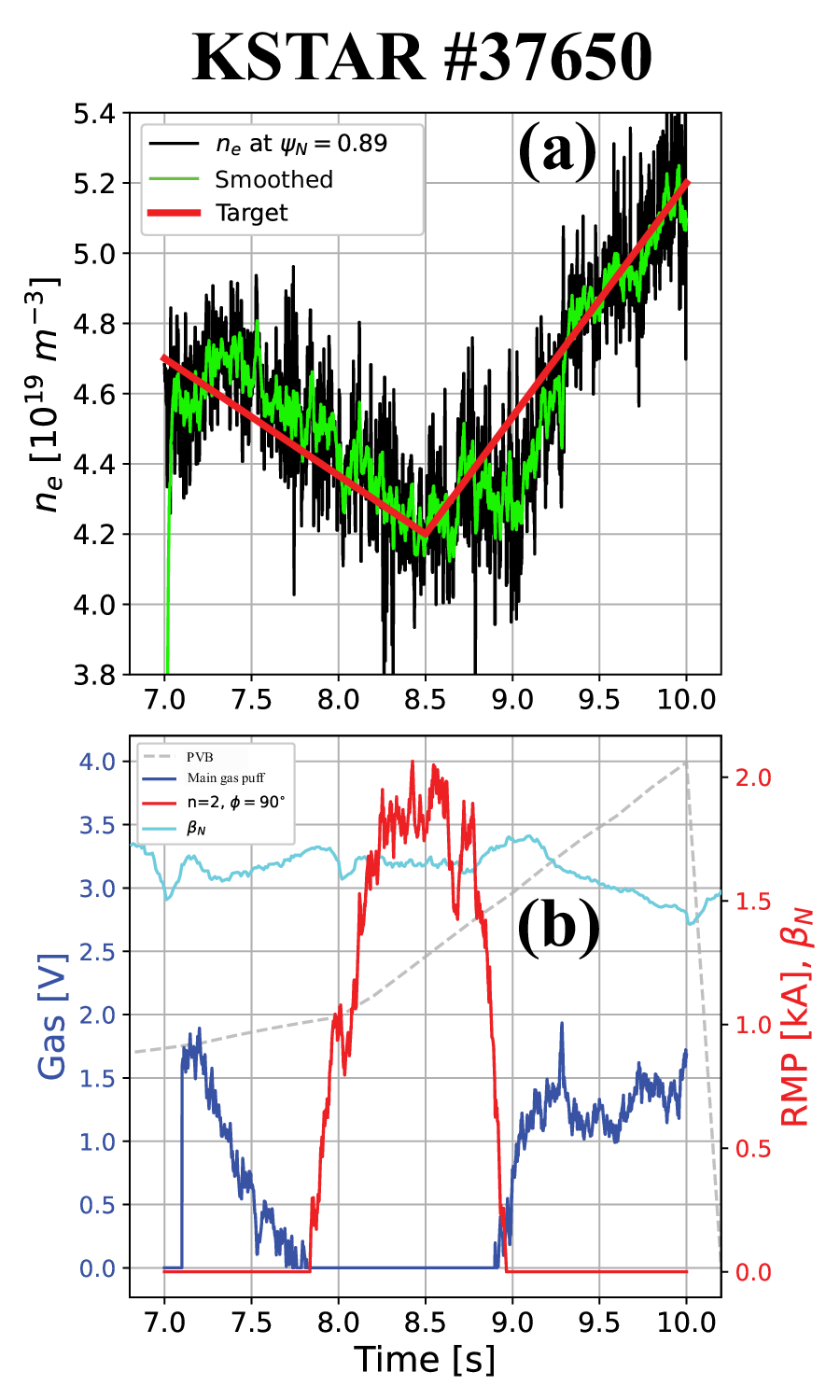}
\caption{Experimental result of controlling $n_\trm{e}$ at $\psi_\trm{N}=0.89$ using n=2, $\phi_{\trm{TM}}=\phi_{\trm{MB}}=90^{\circ}$ RMP and main gas puff as actuators for KSTAR $\#$37650. $\boldsymbol{(a)}$ The black line shows $n_\trm{e}$ at $\psi_\trm{N}=0.89$ reconstructed by a neural network in real-time, and the lime line is the smoothed $n_\trm{e}$ where the smoothing is done by using Eq. \ref{eq:LPF}. The red line represents the control target. $\boldsymbol{(b)}$ The red line shows the RMP current, and the blue line is the voltage applied to the main gas puff. The cyan line represents the $\beta_N$ value calculated by EFIT04 (with linear drift corrected magnetics). Lastly, the gray dashed line shows the voltage applied to the PVB with $D_2$ gas. The controller was turned on at \SI{7}{\second} and turned off at \SI{10}{\second}. As shown in the figures, the controller reacts to minimize the difference between the target and the $n_e$, and the $\beta_N$ value drops when we apply the RMP. The RMP and main gas puff are almost mutually exclusive, as illustrated in the Fig. \ref{fig:err}.}
\label{fig:combo}
\end{center}
\end{figure}

The experimental result of using both actuators to control pedestal-top $n_\trm{e}$ at $\psi_\trm{N}=0.89$ is illustrated in Figure. \ref{fig:combo}. The control experiment was conducted from \SI{7}{\second} to \SI{10}{\second} of KSTAR $\#$37650 with $I_\trm{P}=0.7$MA and $B_\trm{T}=2.6$T. $K_\trm{p}$ and $K_\trm{I}$ for both actuators are given in the Sec. \ref{subsec:PI}. The control target initially decreased linearly, then increased. In this experiment, we set the target more aggressively than in the Figure. \ref{fig:shin} to see if the integrated controller can achieve a more dynamic target. To avoid the saturation of gas input due to the zero initialization of density, the gas command was sent to the PCS after \SI{0.1}{\second} of the start of the controller. Since it was a high-$I_p$ experiment with 0.7 MA, we decided to use a less disruptive $n=2, \ \phi=90^{\circ}$ RMP. We observed less overshoot of the density for RMP compared to the result in the Figure. \ref{fig:shin} since the $n=2$ RMP has a weaker response to the plasmas. The experiment was conducted under the injection of PVB with $D_2$ gas in a feed-forward manner due to experimental constraints. The absolute percentage errors between the $n_\trm{e}$ at $\psi_\trm{N}=0.89$ and the target have a \SI{1.64}{\percent} median and a \SI{2.20}{\percent} average value in KSTAR $\#$37650 during when the controller was on. By using both actuators, the density could follow a more dynamic target compared to the one in the Fig. \ref{fig:shin}. Through the experiments, we found that the gain sets we found by using a simple pole placement logic are generalizable to plasmas with different physical parameters, such as the plasma currents and different actuator configurations. The system identification was done for n=1, $\phi_{\trm{TM}}=\phi_{\trm{MB}}=90^{\circ}$ RMP and PVB with $D_2$ gas but it was also functional for n=2, $\phi_{\trm{TM}}=\phi_{\trm{MB}}=90^{\circ}$ RMP and main gas puff actuators.

\section{Summary and conclusion}
\label{sec:conclusion}
Through the KSTAR 2024-2025 experimental campaign, we could implement a pedestal-top $n_e$ controller that can control $n_\trm{e}$ at $\psi_\trm{N}=0.89$ by using RMP and main gas puff exclusively to follow the target. To get the control target, a real-time $n_\trm{e}$ profile reconstruction algorithm was implemented in KSTAR PCS by using both EFIT and TCI information. The dynamics of plasmas to the actuators was simplified to be a first-order ODE, and the system identification was conducted to find the coefficients $K$ and $T$ of the ODE for both actuators. The transfer function of plasmas could be estimated from the coefficients. A proportional-integration controller was adopted for the experiment, and the pole of the closed-loop transfer function was located at the same location as the pole of the plasma's transfer function. The designed controller enables the $n_\trm{e}$ at $\psi_\trm{N}=0.89$ follow the dynamic target by using only n=1, $\phi_{\trm{TM}} = \phi_{\trm{MB}} = 90^{\circ}$ RMP in Figure. \ref{fig:shin}. To make a complete controller that can follow more dynamic targets, we added main gas puff as another actuator, illustrated in the Figure. \ref{fig:combo}.

We could verify that the pedestal-top $n_e$ can be controlled to follow the dynamic target, which has changing gradients, by using both RMP and gas. Since our controller can either increase or decrease the density dynamically by using both RMP and main gas puff, it allows us to scan the density level within a shot. By doing so, the controller will save costly experimental run time when it is necessary to scan the pedestal-top density under various physical phenomena or new experimental scenarios. The controller can also be combined with the ELM controller or detachment controller to find the window of the pedestal-top $n_{\mathrm{e}}$ for both states. As a next step, we can add other actuators, such as pellet or SMBI (Supersonic Molecular Beam Injection), to control the core and edge region simultaneously.

\section*{Acknowledgement}

This work was supported by the U.S. Department of Energy, Office of Fusion Energy Sciences, under awards DE-SC0024527, DE-SC0020413, DE-SC0015480, DE-SC0021968, and DE-FC02-04ER54698. Additional support was provided by NT-Tao under award NT-Tao-10015925. Support was also provided by the Korean Ministry of Science and ICT under KFE and international R$\&$D Programs (KFE-EN 2503-01).

The computational work and model development were largely carried out using the Princeton Research Computing resources at Princeton University, a consortium led by the Princeton Institute for Computational Science and Engineering (PICSciE) and the Office of Information Technology's Research Computing group.

We also gratefully acknowledge Changmin Shin, JungHoo Hwang, Minsoo Cha, Boseong Kim, and Sanghee Hahn for facilitating data collection during the KSTAR experimental campaign, which was essential for the analysis presented in this paper. We also thank Hyung-Ho Lee, Gun-Young Park, Young-Seok Park, Jaehyun Lee, Giwook Shin, Minwoo Kim, and Yong-Su Na for their insightful discussions regarding experimental planning and the 3D magnetic field configurations.

The authors used Grammarly and OpenAI's ChatGPT to support language editing and revision of this manuscript.

\appendix
\section{Terminologies in control theory}
\label{apx:sys_id}
A system which we want to control is called the plant \cite{book:Rowley} and has input $u(t)$ and output $y(t)$. The plant's response to the delta function input is called the impulse response $h(t)$. For an example of a motor, $u(t)$ would be the voltage we apply, and $y(t)$ would be the angular velocity of the motor. Assuming the system is linear time-invariant (LTI), the input $u(t)$ can be interpreted as a superposition of delta functions with specific amplitudes. Therefore, the output $y(t)$ can be expressed as a convolution in time between the impulse response $h(t)$ and input $u(t)$ \cite{book:Rowley}.

\begin{align}
y(t) &= u(t) * h(t) \nonumber
\\ &= \int^t_0 u(\tau) \cdot h(t-\tau) d\tau.
\label{eq:convolution}
\end{align}

By changing the domain of Eq. \ref{eq:convolution} into the Laplace domain, the convolution can be changed into simple multiplication of the Laplace transformed functions $U(s)$ and $H(s)$.

\begin{align}
Y(s) &= U(s) \cdot H(s), \nonumber
\\ H(s) &= \frac{Y(s)}{U(s)}.
\label{eq:transfer}
\end{align}

$H(s)$ is called the transfer function, which is the frequency response of the plant with the given input. The poles are the points that make the denominator of the transfer function zero. If the real parts of the poles are positive, the system is unstable, and vice versa. System identification is finding the transfer function $H(s)$, so that we can design a controller that changes the dynamics of the plant to satisfy our control requirements, such as overshoot, rising time, settling time, and stability. 

If we model the governing equation of the dynamics between $u(t)$ and $y(t)$, we can find the transfer function that best fits our assumed model. For our experiment, we assumed that the dynamics can be expressed by a first-order ordinary differential equation (ODE). Adopting a simple model is the principle of Occam's razor: selecting a model as simple as possible with the given information \cite{book:murphy}.

\begin{align}
&T \frac{dy}{dt} + y = Ku(t), \label{eq:dynamics}
\\ \trm{where } &T \in \mathbb{R} \trm{ and } K \in \mathbb{R}. \nonumber
\end{align}

By applying Laplace transform to Eq. \ref{eq:dynamics}, 

\begin{align}
sTY(s) + Y(s) = KU(s).
\label{eq:dynamics_laplace}
\end{align}

We can get the transfer function $H(s)$ from the Eq. \ref{eq:dynamics_laplace}, which is

\begin{align}
H(s) = \frac{Y(s)}{U(s)} = \frac{K}{sT+1}.
\label{eq:transfer_laplace}
\end{align}

To return to the time domain where we conduct the convolution between the input $u(t)$ and the impulse response $h(t)$, we apply the inverse Laplace transform to Eq. \ref{eq:transfer_laplace}.

\begin{align}
h(t) = \frac{K}{T} e^{-t/T}
\label{eq:impulse}
\end{align}

\section*{References}
\bibliographystyle{unsrt}
\bibliography{ne_control}

\begin{thebibliography}{10}

\bibitem{paper:wagner1982regime}
Fritz Wagner, G~Becker, K~Behringer, D~Campbell, A~Eberhagen, W~Engelhardt, G~Fussmann, O~Gehre, J~Gernhardt, G~v Gierke, et~al.
\newblock Regime of improved confinement and high beta in neutral-beam-heated divertor discharges of the asdex tokamak.
\newblock {\em Physical Review Letters}, 49(19):1408, 1982.

\bibitem{paper:wagner1984importance}
F~Wagner, M~Keilhacker, ASDEX Team, NI~Team, et~al.
\newblock Importance of the divertor configuration for attaining the h-regime in asdex.
\newblock {\em Journal of Nuclear Materials}, 121:103--113, 1984.

\bibitem{paper:gohil1988study}
P~Gohil, M~Ali Mahdavi, L~Lao, KH~Burrell, MS~Chu, JC~DeBoo, CL~Hsieh, N~Ohyabu, RT~Snider, RD~Stambaugh, et~al.
\newblock Study of giant edge-localized modes in diii-d and comparison with ballooning theory.
\newblock {\em Physical review letters}, 61(14):1603, 1988.

\bibitem{paper:zohm1992studies}
Hartmut Zohm, F~Wagner, M~Endler, J~Gernhardt, E~Holzhauer, W~Kerner, and V~Mertens.
\newblock Studies of edge localized modes on asdex.
\newblock {\em Nuclear fusion}, 32(3):489, 1992.

\bibitem{paper:evans}
Todd~E Evans, Richard~A Moyer, Keith~H Burrell, Max~E Fenstermacher, Ilon Joseph, Anthony~W Leonard, Thomas~H Osborne, Gary~D Porter, Michael~J Schaffer, Philip~B Snyder, et~al.
\newblock Edge stability and transport control with resonant magnetic perturbations in collisionless tokamak plasmas.
\newblock {\em nature physics}, 2(6):419--423, 2006.

\bibitem{paper:kim2020nonlinear}
SK~Kim, S~Pamela, O~Kwon, Marina Becoulet, GTA Huijsmans, Yongkyoon In, M~Hoelzl, JH~Lee, M~Kim, GY~Park, et~al.
\newblock Nonlinear modeling of the effect of n= 2 resonant magnetic field perturbation on peeling-ballooning modes in kstar.
\newblock {\em Nuclear Fusion}, 60(2):026009, 2020.

\bibitem{paper:kim2020pedestal}
Minwoo Kim, J~Lee, WH~Ko, S-H Hahn, Yongkyoon In, YM~Jeon, W~Suttrop, SK~Kim, GY~Park, J-W Juhn, et~al.
\newblock Pedestal electron collisionality and toroidal rotation during elm-crash suppression phase under n= 1 rmp in kstar.
\newblock {\em Physics of Plasmas}, 27(11), 2020.

\bibitem{paper:yang2024tailoring}
SeongMoo Yang, Jong-Kyu Park, YoungMu Jeon, Nikolas~C Logan, Jaehyun Lee, Qiming Hu, JongHa Lee, SangKyeun Kim, Jaewook Kim, Hyungho Lee, et~al.
\newblock Tailoring tokamak error fields to control plasma instabilities and transport.
\newblock {\em Nature Communications}, 15(1):1275, 2024.

\bibitem{paper:kim2024highest}
SangKyeun Kim, Ricardo Shousha, SeongMoo Yang, Qiming Hu, SangHee Hahn, Azarakhsh Jalalvand, J-K Park, Nikolas~Christopher Logan, Andrew~Oakleigh Nelson, Y-S Na, et~al.
\newblock Highest fusion performance without harmful edge energy bursts in tokamak.
\newblock {\em Nature communications}, 15(1):3990, 2024.

\bibitem{paper:shousha2022design}
Ricardo Shousha, Sang~Kyeun Kim, Keith~G Erickson, SH~Hahn, AO~Nelson, Seong~Moo Yang, J-K Park, Josiah Wai, YM~Jeon, JH~Lee, et~al.
\newblock Design and experimental demonstration of feedback adaptive rmp elm controller toward complete long pulse elm suppression on kstar.
\newblock {\em Physics of Plasmas}, 29(3), 2022.

\bibitem{paper:kim2022optimization}
SangKyeun Kim, Ricardo Shousha, SH~Hahn, Andrew~O Nelson, J~Wai, Seong~Moo Yang, J-K Park, R~Nazikian, Nikolas~C Logan, YM~Jeon, et~al.
\newblock Optimization of 3d controlled elm-free state with recovered global confinement for kstar with n= 1 resonant magnetic field perturbation.
\newblock {\em Nuclear Fusion}, 62(2):026043, 2022.

\bibitem{paper:shousha2023thesis}
Ricardo Shousha.
\newblock {\em Real-time kinetic profile reconstruction and Adaptive ELM Control on the DIII-D and KSTAR Tokamaks}.
\newblock PhD thesis, ProQuest Dissertations Publishing, 2023.

\bibitem{paper:matthews1995plasma}
GF~Matthews.
\newblock Plasma detachment from divertor targets and limiters.
\newblock {\em Journal of nuclear materials}, 220:104--116, 1995.

\bibitem{paper:loarte1998plasma}
A~Loarte, RD~Monk, JR~Mart{\'\i}n-Sol{\'\i}s, DJ~Campbell, AV~Chankin, S~Clement, SJ~Davies, J~Ehrenberg, SK~Erents, HY~Guo, et~al.
\newblock Plasma detachment in jet mark i divertor experiments.
\newblock {\em Nuclear Fusion}, 38(3):331, 1998.

\bibitem{paper:loarte2001effects}
Alberto Loarte.
\newblock Effects of divertor geometry on tokamak plasmas.
\newblock {\em Plasma Physics and Controlled Fusion}, 43(6):R183, 2001.

\bibitem{abstract:hu2023integration}
Qiming Hu, Huiqian Wang, David Eldon, Shuai Gu, Heinke Frerichs, Filippo Scotti, Robert Wilcox, Alessandro Bortolon, Lennard Ceelen, Florian Effenberg, et~al.
\newblock Integration of rmp elm control with divertor detachment in the diii-d tokamak.
\newblock In {\em APS Division of Plasma Physics Meeting Abstracts}, volume 2023, pages BI02--003, 2023.

\bibitem{paper:laggner2020real}
FM~Laggner, D~Eldon, AO~Nelson, C~Paz-Soldan, A~Bortolon, TE~Evans, ME~Fenstermacher, BA~Grierson, Q~Hu, DA~Humphreys, et~al.
\newblock Real-time pedestal optimization and elm control with 3d fields and gas flows on diii-d.
\newblock {\em Nuclear Fusion}, 60(7):076004, 2020.

\bibitem{paper:hawryluk2015control}
Richard~J Hawryluk, Nicholas~W Eidietis, Brian~A Grierson, Alan~W Hyatt, Egemen Kolemen, Nikolas~C Logan, R~Nazikian, C~Paz-Soldan, Wayne~M Solomon, and S~Wolfe.
\newblock Control of plasma stored energy for burn control using diii-d in-vessel coils.
\newblock {\em Nuclear Fusion}, 55(5):053001, 2015.

\bibitem{paper:EFIT}
Lang~L Lao, H~St John, RD~Stambaugh, AG~Kellman, and W~Pfeiffer.
\newblock Reconstruction of current profile parameters and plasma shapes in tokamaks.
\newblock {\em Nuclear fusion}, 25(11):1611, 1985.

\bibitem{paper:CAKE}
ZA~Xing, D~Eldon, AO~Nelson, MA~Roelofs, WJ~Eggert, O~Izacard, AS~Glasser, NC~Logan, O~Meneghini, SP~Smith, et~al.
\newblock Cake: consistent automatic kinetic equilibrium reconstruction.
\newblock {\em Fusion Engineering and Design}, 163:112163, 2021.

\bibitem{paper:Li2013}
GQ~Li, QL~Ren, JP~Qian, LL~Lao, SY~Ding, YJ~Chen, ZX~Liu, B~Lu, and Q~Zang.
\newblock Kinetic equilibrium reconstruction on east tokamak.
\newblock {\em Plasma Physics and Controlled Fusion}, 55(12):125008, 2013.

\bibitem{paper:jiang2021}
Yanzheng Jiang, SA~Sabbagh, YS~Park, JW~Berkery, JH~Ahn, JD~Riquezes, JG~Bak, WH~Ko, J~Ko, JH~Lee, et~al.
\newblock Kinetic equilibrium reconstruction and the impact on stability analysis of kstar plasmas.
\newblock {\em Nuclear Fusion}, 61(11):116033, 2021.

\bibitem{paper:juhn2021multi}
June-Woo Juhn, KC~Lee, TG~Lee, HM~Wi, YS~Kim, SH~Hahn, and YU~Nam.
\newblock Multi-chord ir--visible two-color interferometer on kstar.
\newblock {\em Review of Scientific Instruments}, 92(4), 2021.

\bibitem{paper:shousha2023machine}
Ricardo Shousha, Jaemin Seo, Keith Erickson, Zichuan Xing, SangKyeun Kim, Joseph Abbate, and Egemen Kolemen.
\newblock Machine learning-based real-time kinetic profile reconstruction in diii-d.
\newblock {\em Nuclear Fusion}, 64(2):026006, 2023.

\bibitem{paper:rothstein2025torbeamnn}
Andrew Rothstein, Minseok Kim, Minho Woo, Minsoo Cha, Cheolsik Byun, Sangkyeun Kim, Keith Erickson, Youngho Lee, Josh Josephy-Zack, Jalal Butt, et~al.
\newblock Torbeamnn: machine learning-based steering of ech mirrors on kstar.
\newblock {\em Plasma Physics and Controlled Fusion}, 67(5):055036, 2025.

\bibitem{paper:boyer2021prediction}
Mark~D Boyer and Jason Chadwick.
\newblock Prediction of electron density and pressure profile shapes on nstx-u using neural networks.
\newblock {\em Nuclear Fusion}, 61(4):046024, 2021.

\bibitem{paper:joung2019deepNN}
Semin Joung, Jaewook Kim, Sehyun Kwak, JG~Bak, SG~Lee, HS~Han, HS~Kim, Geunho Lee, Daeho Kwon, and Y-C Ghim.
\newblock Deep neural network grad--shafranov solver constrained with measured magnetic signals.
\newblock {\em Nuclear Fusion}, 60(1):016034, 2019.

\bibitem{paper:joung2023gs}
Semin Joung, Y-C Ghim, Jaewook Kim, Sehyun Kwak, Daeho Kwon, C~Sung, D~Kim, Hyun-Seok Kim, JG~Bak, and SW~Yoon.
\newblock Gs-deepnet: mastering tokamak plasma equilibria with deep neural networks and the grad--shafranov equation.
\newblock {\em Scientific Reports}, 13(1):15799, 2023.

\bibitem{book:bishop}
Christopher M.Bishop.
\newblock {\em Pattern Recognition and Machine Learning}.
\newblock Springer, 2006.

\bibitem{book:Astrom}
Karl~J \AA{}str\"om and Tore H\"agglund.
\newblock {\em PID controllers}.
\newblock Instrument Society of America, 2nd edition, 1995.

\bibitem{book:Rowley}
Clarence~W Rowley.
\newblock {\em Introduction to Feedback Control}.
\newblock Clarence W Rowley, 2024.

\bibitem{paper:kim2009IVCCdesign}
HK~Kim, HL~Yang, GH~Kim, Jin-Yong Kim, Hogun Jhang, JS~Bak, and GS~Lee.
\newblock Design features of the kstar in-vessel control coils.
\newblock {\em Fusion Engineering and Design}, 84(2-6):1029--1032, 2009.

\bibitem{paper:kim2011IVCCfabrication}
HK~Kim, KS~Lee, HL~Yang, JR~Last, E~Bertolini, KM~Kim, EN~Bang, HT~Kim, YM~Jeon, and M~Kwon.
\newblock Fabrication and installation of kstar in-vessel control coils.
\newblock {\em Fusion engineering and design}, 86(9-11):1975--1979, 2011.

\bibitem{kim2013gas}
Young~Ok Kim, Jae~In Song, Kwang~Pyo Kim, Yong Chu, Kap~Rai Park, Hong~Tack Kim, Hak~Kun Kim, Kun~Su Lee, and Yang~Mo Kim.
\newblock Control and operation of the gas injection systems for kstar tokamak.
\newblock {\em Fusion Engineering and Design}, 88(6-8):1132--1136, 2013.

\bibitem{paper:fitting}
PB~Snyder, HR~Wilson, JR~Ferron, LL~Lao, AW~Leonard, D~Mossessian, M~Murakami, TH~Osborne, AD~Turnbull, and XQ~Xu.
\newblock Elms and constraints on the h-mode pedestal: peeling--ballooning stability calculation and comparison with experiment.
\newblock {\em Nuclear fusion}, 44(2):320, 2004.

\bibitem{book:Rasmussen}
Carl~Edward Rasmussen and Christopher K.~I. Williams.
\newblock {\em Gaussian Processes for Machine Learning}.
\newblock The MIT Press, 2006.

\bibitem{paper:ms_GPR}
Minseok Kim, WH~Ko, Sehyun Kwak, Semin Joung, Wonjun Lee, B~Kim, D~Kim, JH~Lee, Choongki Sung, Yong-Su Na, et~al.
\newblock Kinetic profile inference with outlier detection using support vector machine regression and gaussian process regression.
\newblock {\em Nuclear Fusion}, 64(10):106052, 2024.

\bibitem{paper:kwak}
Sehyun Kwak, Jakob Svensson, S~Bozhenkov, Joanne Flanagan, Mark Kempenaars, Alexandru Boboc, Y-C Ghim, and JET Contributors.
\newblock Bayesian modelling of thomson scattering and multichannel interferometer diagnostics using gaussian processes.
\newblock {\em Nuclear Fusion}, 60(4):046009, 2020.

\bibitem{paper:juhn}
June-Woo Juhn, KC~Lee, TG~Lee, HM~Wi, YS~Kim, SH~Hahn, and YU~Nam.
\newblock Multi-chord ir--visible two-color interferometer on kstar.
\newblock {\em Review of Scientific Instruments}, 92(4), 2021.

\bibitem{paper:kolemen2010strike}
Egemen Kolemen, DA~Gates, Clarence~Worth Rowley, N~Jeremy Kasdin, J~Kallman, S~Gerhardt, V~Soukhanovskii, and D~Mueller.
\newblock Strike point control for the national spherical torus experiment (nstx).
\newblock {\em Nuclear fusion}, 50(10):105010, 2010.

\bibitem{paper:park2013investigation}
YS~Park, SA~Sabbagh, JM~Bialek, JW~Berkery, SG~Lee, WH~Ko, JG~Bak, YM~Jeon, JK~Park, J~Kim, et~al.
\newblock Investigation of mhd instabilities and control in kstar preparing for high beta operation.
\newblock {\em Nuclear Fusion}, 53(8):083029, 2013.

\bibitem{thesis:JWJuhn2013study}
June-Woo Juhn.
\newblock {\em Study on Global Particle Balance Model for Plasma Density Feedback Control in KSTAR}.
\newblock PhD thesis, Seoul National University, 2013.

\bibitem{paper:fahmy2016multivariable}
Rania~A Fahmy, Ragia~I Badr, and Farouk~A Rahman.
\newblock Multivariable online adaptive pid controller for plasma current, shape, and position in tokamaks.
\newblock {\em Journal of Fusion Energy}, 35:831--840, 2016.

\bibitem{paper:ravensbergen2017density}
Timo Ravensbergen, Peter~C de~Vries, Federico Felici, Thomas~Cornelis Blanken, Remy Nouailletas, and L~Zabeo.
\newblock Density control in iter: an iterative learning control and robust control approach.
\newblock {\em Nuclear Fusion}, 58(1):016048, 2017.

\bibitem{paper:blanken2019model}
TC~Blanken, F~Felici, C~Galperti, O~Kudl{\'a}{\v{c}}ek, F~Janky, A~Mlynek, L~Giannone, PT~Lang, W~Treutterer, WPMH Heemels, et~al.
\newblock Model-based real-time plasma electron density profile estimation and control on asdex upgrade and tcv.
\newblock {\em Fusion Engineering and Design}, 147:111211, 2019.

\bibitem{book:murphy}
Kevin~P. Murphy.
\newblock {\em Machine Learning: A Probabilistic Perspective}.
\newblock The MIT Press, 2012.

\end{thebibliography}

\end{document}